# Hardware Security Evaluation of MAX 10 FPGA

Feasibility Study of Intel® MAX 10 devices for compliance to MODH security level

Sergei Skorobogatov

Dept of Computer Science and Technology
University of Cambridge
Cambridge, UK
sps32@cam.ac.uk

*Abstract*—With the ubiquity of IoT devices there is a growing demand for confidentiality and integrity of data. Solutions based on reconfigurable logic (CPLD or FPGA) have certain advantages over ASIC and MCU/SoC alternatives. Programmable logic devices are ideal for both confidentiality and upgradability purposes. In this context the hardware security aspects of CPLD/FPGA devices are paramount. This paper shows preliminary evaluation of hardware security in Intel® MAX 10 devices. These FPGAs are one of the most suitable candidates for applications demanding extensive features and high level of security. Their strong and week security aspects are revealed and some recommendations are suggested to counter possible security vulnerabilities in real designs. This is a feasibility study paper. Its purpose is to highlight the most vulnerable areas to attacks aimed at data extraction and reverse engineering. That way further investigations could be performed on specific areas of concern.

## I. Introduction

Modern systems with wireless connectivity and Internet access require robust hardware security to prevent all sorts of attacks [1,2,3]. In order to comply with these demands modern semiconductor chips are designed with hardware security in mind.

Recent scandal with modified server motherboards manufactured by Supermicro [4,5] forced developers to improve the security through the monitoring of system booting process. Solutions based on trusted components are the best choice. However, reprogrammable components such as SoCs, CPLDs or FPGAs offer more flexibility and better control with the ability to later upgrades in the field. In modern motherboards BIOS is usually located in SPI Flash susceptible to modification attacks [6,7]. Given the high speed of SPI bus only solutions based on CPLD or FPGA will be operating in real time. Moreover, the hardware security of such devices is usually higher given the fact that any reverse engineering will involve significantly more efforts. This is because unlike CPU that executes code sequentially, CPLD and FPGA perform operations in parallel based on logic implementation and state machines. These days the border between CPLD and FPGA is blended. Some latest CPLDs contain special functional blocks and more logic elements than small FPGAs. However, all CPLDs can operate immediately after power-up, while most FPGAs require external bitstream to be uploaded, usually via slow serial interface. Certain FPGA families have on-chip or in-package Flash to hold the configuration image. Some FPGAs offer bitstream encryption to prevent cloning and reverse engineering, but some security vulnerabilities were still found in them [8,9,10]. Most CPLD and FPGA devices are configured using JTAG interface [11]. However, security flaws were still found in specific JTAG designs that allowed readback of the on-chip configuration bitstream [12].

TABLE I. CPLD AND FPGA FAMILIES WITH ON-CHIP CONFIGURATION

| Devices | Logic Elements | User Flash kbit | NVM Images | Min Size mm$^2$ | Min Cost USD | Security | CPU core |
|---|---|---|---|---|---|---|---|
| Microsemi ProAsic3 130nm | 330 – 11k LUT3 | 1 | 1 | 6×6 | 3 | Lock Verify AES | ARM M1 soft |
| Microsemi Smart Fusion2 65nm | 6k – 146k LUT4 | 1024 – 4096 | 1 | 11×11 | 8 | Lock Verify CRC AES | ARM M3 hard |
| Microsemi PolarFire 28nm | 100k – 500k LUT4 | 297 – 513 | 1 | 11×11 | 110 | Lock Verify CRC AES | ARM M4 hard |
| Lattice XP2 90nm | 5k – 40k LUT4 | 0 | 1 | 8×8 | 3 | Lock | none |
| Lattice MachXO2 65nm | 256 – 6864 LUT4 | 0 – 256 | 1 | 2.5×2.5 | 2 | Verify | Mico8 soft |
| Lattice MachXO3 65nm | 640 – 9400 LUT4 | 64 – 448 | 1 | 2.5×2.5 | 3 | Lock Verify | Mico32 soft |
| Lattice MachXO3D 65nm | 4300 – 9400 LUT4 | 367 – 2693 | 2 | 10×10 | 20 | Lock Verify | Mico32 soft |
| Intel MAX 10 55nm | 2k – 50k LUT4 | 96 – 5888 | 2 | 3×3 | 2.50 | Lock Verify CRC AES | NIOSII soft |

Table 1 outlines the modern reprogrammable live-on-power-up FPGA families. Only latest devices fabricated with 65nm or smaller process offer decent amount of user Flash memory. Intel MAX 10 devices offer storage for two configuration images thus making the update process much safer against any interruptions. A golden image can always be stored on-chip, that way any failures during the update process can be handled efficiently without the need of establishing a secure reprogramming environment. Also, if bugs are found in the new firmware it can be quickly rolled back using the recovery function implemented in the golden image.

Active monitoring devices used for security applications such as server motherboards will likely have to meet certain requirements, for example, NIST Root-of-Trust standard [13]. Given the extensive features of Intel MAX 10 FPGAs (dual boot from on-chip flash, user flash, security lock) they seem to be the most likely candidates for this purpose [14,15]. They offer good balance between the number of elements, user Flash memory size, package size, and cost. However, for the end users the security is paramount. Therefore, this paper is focused on hardware security evaluation of some MAX 10 devices.

The paper is organised as follows. Section 2 gives introduction to hardware security, attack technologies and differences between CPLD/FPGA and MCU/SoC solutions. Section 3 describes experimental setups for evaluation against various attacks. Section 4 shows the results, and Section 5 discusses them. Section 6 outlines future work, and Section 7 concludes the paper.

## II. BACKGROUND

Hardware Security helps with designing a secure system by combining the knowledge of existing attack technologies with state-of-the-art defence technologies. Without awareness about modern attack methods it would be impossible to design a secure system. Countermeasures comes in help to assist the design process. However, for economical and convenience reasons it is not always possible to incorporate all countermeasures. Hence, the job of the well educated hardware designer is to choose the right solutions.

### A. Hardware Security and attack technologies

The attacks on semiconductor devices can be split into several categories. Non-invasive attacks which are usually low-cost and involve observations of the device operation or manipulations with external signals. They require only moderately sophisticated equipment and knowledge to implement. They do not physically harm the chip and often leave no traces. Invasive attacks, in contrast, are expensive and require sophisticated equipment and knowledgeable attackers. However, they offer almost unlimited capabilities to extract information from chips and understand their functionality. These attacks always leave traces and often destroy the chip. Semi-invasive attacks fill the gap between non-invasive and invasive attacks and usually affordable to many attackers. For these attacks the chip needs to be depackaged but the internal structure remains intact. Although these attacks often leave traces, in most cases the chip remains fully operational.

Tools used for carrying out non-invasive attacks are usually available at most electronics engineering labs. These tools involve digital multimeter, IC soldering/desoldering station, universal programmer, oscilloscope, logic analyser, signal generator, power supply, PC and prototyping boards.

Non-invasive attacks can be divided into side-channel attacks (timing [16], power analysis [17], emission analysis [18]), data remanence [19], data mirroring [20], fault injection (glitching [21], bumping [22]) and brute forcing [23].

Tools used for carrying out invasive attacks involve simple chemical lab, high-resolution optical microscope, wire bonding machine, laser cutting system, microprobing station, oscilloscope, logic analyser, signal generator, power supply, scanning electron microscope (SEM) and focus ion beam (FIB) workstation.

Invasive attacks can be divided into sample preparation [24], imaging [25], direct memory extraction [26], reverse engineering [27], microprobing [28], fault injection [29] and chip modification [30].

Tools used for carrying out semi-invasive attacks involve simple chemical lab, high-resolution optical microscope, UV light source, lasers, oscilloscope, logic analyser, signal generator, PC and prototyping boards.

Semi-invasive attacks can be divided into imaging [31], laser scanning [32], optical fault injection [33], optical emission analysis [34] and combined attacks [35].

### B. Advantages of CPLD/FPGA over MCU/SoC solutions

Complex Programmable Logic Device (CPLD) consist from limited number of building blocks – macrocells. Each has a programmable logic expression and a flip-flop with configurable inputs and outputs. These devices are usually relatively small (32 to 512 macrocells), slow (about 100MHz), have limited number of I/O pins (about a hundred), but inexpensive and live on power-up. Their configuration memory is usually based on EEPROM with proprietary programming interface and some security features against readback.

Field Programmable Gate Array (FPGA) consist from large number of logic blocks: LUT, Flip-Flops, SRAM, DSP, I/O interfaces. It has a hierarchy of reconfigurable interconnects. These devices are usually large (thousands of logic elements), fast (hundreds of MHz speed), large number of I/O pins (hundreds to thousands), but expensive and require time to upload the bitstream into configuration SRAM.

These days the gap between CPLD and FPGA is shrinking. Some latest CPLDs contain special functional blocks and more logic elements than some small FPGAs. However, both CPLDs and FPGAs share the same advantage over CPU-based microcontrollers (MCU) and system-on-chip (SoC) devices – they have programmable logic and perform operations in parallel.

### C. Reverse engineering challenges

The reverse engineering is a way to understand how a particular device works. From an attacker point of view reverse engineering allows ultimate insight into the device functionality and understanding how its security protection operates. With scrupulous analysis of the security features some vulnerabilities and weaknesses could be found. This could potentially lead to powerful and inexpensive attack, especially with non-invasive methods. Those methods could be easily replicated and transferred to other attackers. In some cases mass produced low-cost attacking tools could be built [36,37].

Reverse engineering of a design implemented in CPLD or FPGAs is significantly harder than a one implemented in microcontroller or SoC. This is because no such tools as disassembler exist for logic devices. Therefore, even if the bitstream is successfully extracted it would take time to convert it into a netlist or schematic for further analysis.

III. EXPERIMENTAL SETUP

For experiments the following Intel MAX 10 devices were used: dual supply 10M08DCF256, 10M16DAF256, and single supply 10M04SCE144, 10M08SAE144, 10M08SCE144, 10M16SCE144.

A. *Non-invasive attacks*

For the initial non-invasive experiments the Intel MAX 10 Evaluation Kits with 10M08SAE144 and 10M16SCE144 devices were used. Some TQFP144 devices were programmed in the Elnec BeeProg2 universal programmer using PLD-18 adapter. For most experiments the devices were placed into a ZIF socket attached to custom built test board (Figure 1). Then a JTAG cable was connected to either the Altera USB Byte Blaster, modified Elnec PLD-18 adapter (Figure 2) or the custom built control board (Figure 3).

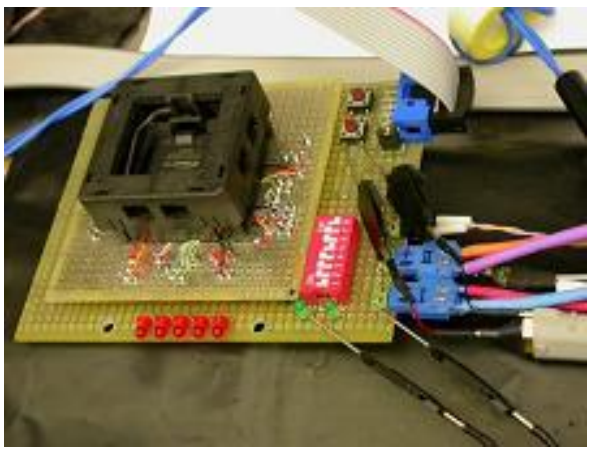

Fig. 1. Test board with BGA256 ZIF socket.

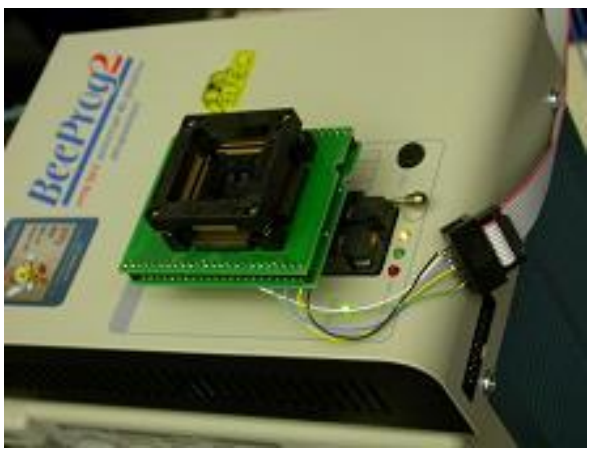

Fig. 2. Elnec BeeProg2 with modified PLD-18 adapter.

The test board had 2.2Ω resistors in $V_{CC}$, $V_{CCA}$ and $V_{IO}$ power supply lines for power analysis experiments using a 1GHz digital storage oscilloscope with differential probes. The control board had an 80MIPS PIC24 microcontroller with implementation of JTAG protocol and was connected to a PC running custom software.

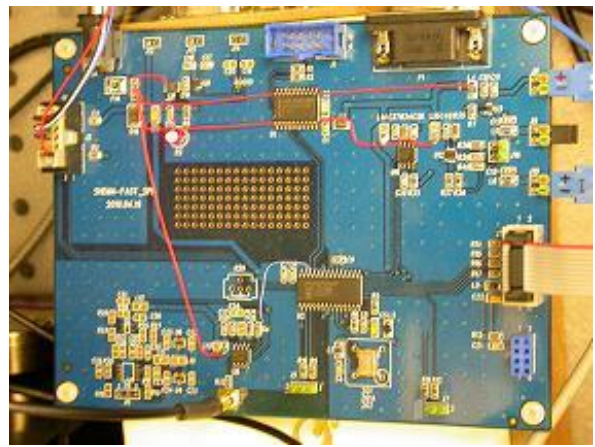

Fig. 3. Custom built control board.

The test board with MAX 10 device was powered by system power supplies with programmable voltages. For power glitching experiments an arbitrary waveform signal generator was used to provide glitch timing. The board had a dedicated switch for connecting pre-selected power supply to the core supply circuit of FPGAs.

For electromagnetic analysis (EMA) an H-probe was used with a 50dB amplifier connected to an oscilloscope. The probe was positioned over different areas on the chip surface using a motorised stage.

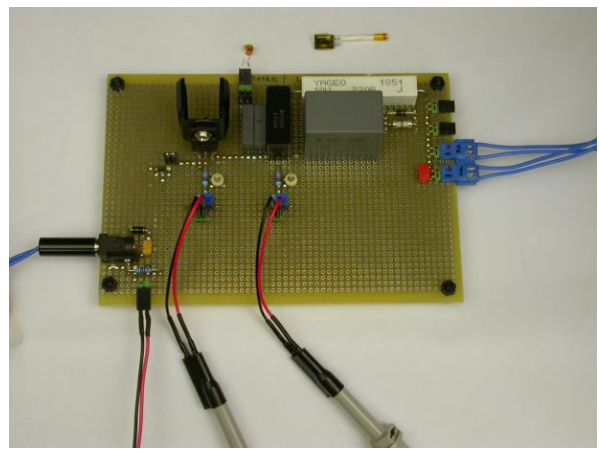

Fig. 4. Custom built board for electromagnetic fault injection.

For electromagnetic (EM) fault injection a special pulse generation board was built (Figure 4). It was supplying a high voltage pulse of up to 500V to a small custom-built coil. The duration of the pulse from 10ns to 100ns was determined by a signal generator with 1ns precision timing. The actual fault was produced by a flyback voltage pulse generated by the coil when it was disconnected by a transistor on the board. It is almost impossible to get rid of this flyback voltage pulse generated by the coil, but it was possible to use it as the main source for fault injection.

For eavesdropping on the JTAG protocol used in MAX 10 a 500MHz logic analyser was used.

Data acquisition and processing were carried out on a PC using Matlab software.

### B. Semi-invasive attacks

For semi-invasive experiments the TQFP144 devices were first opened from the backside before being placed into the test board under optical microscope with IR laser input (Figure 5). The power of the laser was controlled from 1mW to 100mW.

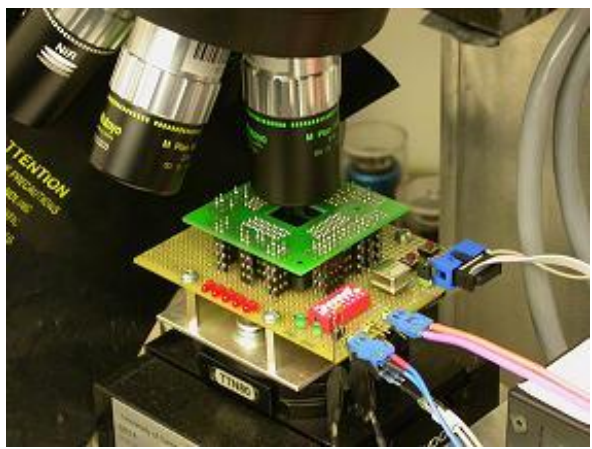

Fig. 5. Test board mounted under microscope.

### C. Invasive attacks

For invasive sample preparation 10M16SAU169 device was first mechanically polished from the balls side down to silicon thickness of 10 microns. Then the remaining silicon was removed with chemical etching.

### IV. RESULTS

The process of creating a design and configuration files was first assessed. For that the Quartus Prime development software was installed and example files from the Intel MAX 10 Evaluation Kit were used. The output of successfully compiled design is in *.sof file. It can be used to directly configure SRAM, but in order to program the configuration Flash it must be converted into *.pof file. During this process several security fuses can be activated, these are Verify Protect, Encrypted POF Only and JTAG Security fuses. For encrypted configuration also *.ekp file containing AES key must be created.

Quartus Programmer tool can use *.pof files to program MAX 10 devices. It can also convert them to *.jam STAPL files [38]. Software of Elnec Universal programmer can convert *.pof files into *.hex files. This simplifies their analysis. However, the same *.hex file can be extracted from a variable in *.jam file.

Small changes in the design (flip 1 bit of data) result in 1-bit change in the main body of *.sof file. However, *.sof file has a 16-byte number at address 008Bh which is unique for each design, design checksum at address 0114h and CRC at the end − all proprietary except checksum. In the *.pof file there is 1-bit difference in the design area and multiple differences in system area, as well as proprietary CRC at the end. The *.jam file incorporates data from *.pof file plus system configuration.

All configuration files have proprietary format. Although the JTAG communication can be eavesdropped using logic analyser, it does not help in understanding the bitstream format. Also, any single-bit change in the data stream results in the non-operational device because of multiple CRC checks.

For 10M08SCE144 device the mapping file created by Quartus software contains the following information:

```
ICB    0x00000000  0x000007FF
UFM    0x00000800  0x0001CFFF
CFM0   0x0001D000  0x0004E7FF (0x0004628F)
EPOF: OFF
Secured JTAG: OFF
Verify protect: OFF
Watchdog value: Not activated
POR: Instant ON
IO Pullup: ON
SPI IO Pullup: ON
Data checksum for this conversion is 0x0266384E
All the addresses in this file are byte addresses
```

Quartus tool generates different *.pof files depending on the settings of fuses, in both system (ICB) and data (CFM0) areas. For Verify Protect fuse:

0030h: 0F,A5,48,6C    01D007h: D2    01D00Ch: F3,0C,59

For Encrypted Bitstream with AES key:

0000h: 16-byte scrambled key 01234...EF ==> 3B7F195D2A6E084C
0028h: 0F,A5,48,6C    01D007h: C2    01D00Ch: E2,0C,98

For Encrypted POF Only fuse:

0014h: 0F,A5,48,6C    01D007h: C3    01D00Ch: F2,0C,58

For Secured JTAG fuse:

001Ch: 0F,A5,48,6C    01D007h: C6    01D00Ch: A7,0C,58

STAPL file analysis is complicated by some obfuscation. Variables have obscured names (A12, V185). Subroutines have meaningless names (L107, L1259). However, IRSCAN, DRSCAN and WAIT commands can be found. What also helps is adding PRINT command for subroutine name, JTAG command, size and data. Then all the communication information from dialog screen can be copied into a file for further analysis.

Quartus Programmer works differently from Universal programmer. In Quartus software even if Verify Protect fuse is set the verification is allowed, but Examine (readback) fails above 1D000h for 10M08SCE144 chip. Encrypted POF Only fuse prevents Program, Verify and Blank Check. In Universal programmer Verify Protect fuse prevents verification and reading above 1D000h. But Encrypted POF Only has no effect on security. This is caused by the preload of SRAM in Quartus tool and the use of JTAG user mode. However, this exposes undocumented security flaw in MAX 10 devices. If only Verify Protect fuse is set it does not protect the configuration memory. Neither Encrypted POF Only fuse on its own protects anything. It is only the combination of both Verify Protect and Encrypted POF Only fuses that protects the design.

When the bitstream is encrypted it is not straightforward to decrypt it even if the key is known. The mode of AES operation is not documented. Moreover, the sequence of input bytes and initialisation vector are also unknown. It is very likely that some reverse engineering of Quartus tools might be required to extract this information.

*A. Non-invasive attacks*

Power analysis reveals some information about internal device operation. Figure 6 shows the boot process from internal Flash. If the bitstream is encrypted with different keys there is a noticeable difference in the power trace (Figure 7).

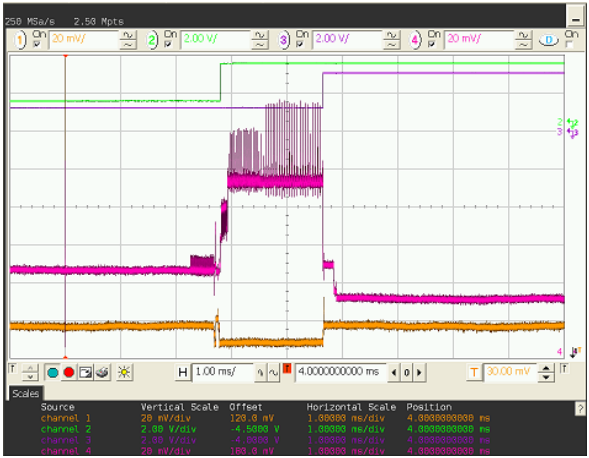

Fig. 6. Boot process from internal Flash.

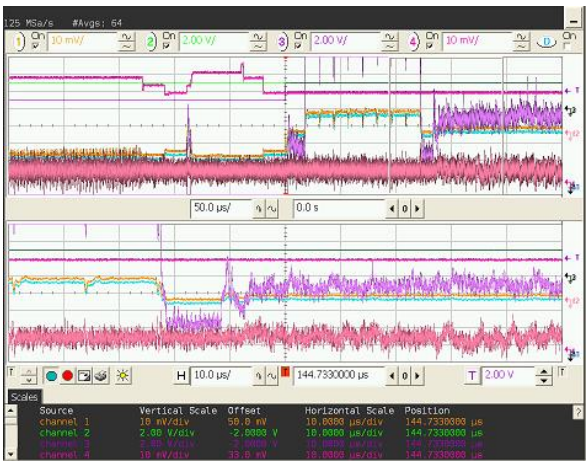

Fig. 7. Difference in power trace between correct and incorrect keys.

When the encrypted bitstream is corrupted this results in termination of the boot process. However, the differences between the last AES decryption operation is clearly visible (Figure 8).

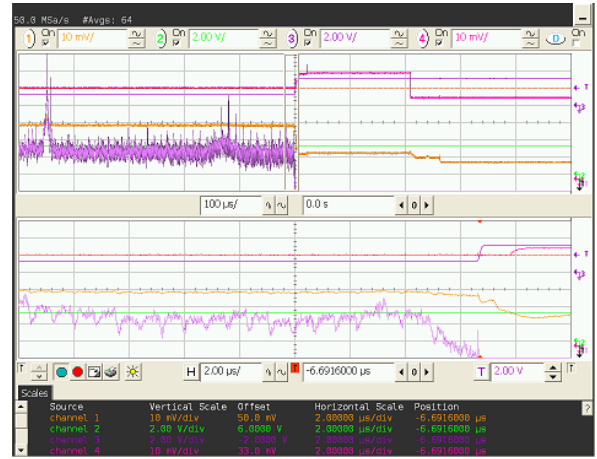

Fig. 8. Difference in power traces for corrupted bitstream.

There was no success with EMA experiments. First, the signal-to-noise ratio was significantly smaller compared to the power analysis even when the H-probe coil was placed above the most contributing area. Second, the contribution of the internal clock signal to the trace was an order of magnitude larger than to the power analysis trace.

TABLE II. POWER GLITCHING RESULTS FOR SINGLE SUPPLY MAX 10

| Device | Glitch parameters | | | |
|---|---|---|---|---|
| | 1.5V 5µs | 1.45V 4µs | 1.4V 4µs | 1.3V 3.5µs |
| 10M16SCE144 | 9 | 1706 | 1860 | 17 |

TABLE III. POWER GLITCHING RESULTS FOR DUAL SUPPLY MAX 10

| Device | Glitch parameters | | | |
|---|---|---|---|---|
| | 0.6V 1.2µs | 0.4V 0.7µs | 0.3V 0.5µs | 0.2V 0.4µs |
| 10M16DAF256 | 241 | 650 | 13491 | 9954 |

Some MAX 10 devices were tested for sensitivity to power glitching. For that a signal generator was used to supply the VCC of the device with unstable power. This was carried out during the Examine operation (reading the Flash contents) in the Quartus Programmer. It was observed that some power glitches passed through the internal detection circuit without triggering the device restart with boot process. This resulted in some data corruption. The results for single supply MAX 10 device is presented in Table 2. The glitch parameter corresponds to the reduction of the core supply voltage from the nominal value of 3.0V. Table 3 presents the results for dual supply MAX 10 device. The glitch parameter corresponds to the reduction of the core supply voltage from the nominal value of 1.2V. The value in the table corresponds to the number of incorrect reads from the whole Flash memory of the device during JTAG read operation. It can be observed that dual supply devices are more sensitive to power glitching due to the direct connection of the core supply grid with the pins. The lower the peak voltage of the glitch the higher is the error rate. However, if the peak voltage is too low it triggers the reset of the device.

The example of a successful power glitch is presented in Figure 9. It was causing the wrong data to be fetched from the on-chip Flash memory thus potentially allowing an attacker to circumvent the security protection fuses.

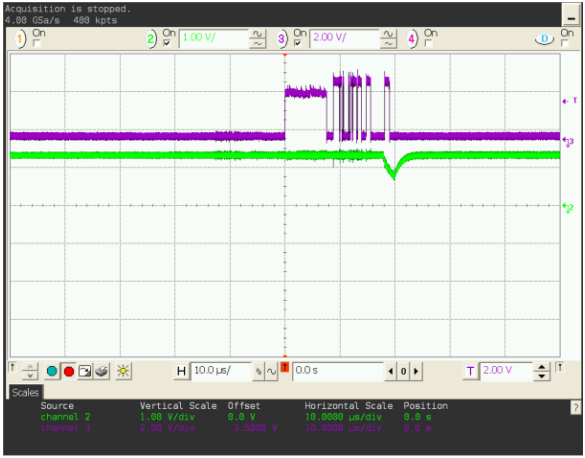

Fig. 9. Example of a successful power glitch.

The results of testing MAX 10 devices against electromagnetic fault injection are presented in Table 4 for single supply device and in Table 5 for dual supply device. Both devices showed good sensitivity to the glitching during readout of the Flash memory. As with the power glitching if the peak voltage of the pulse was too high it caused the device to malfunction. The example of a typical voltage on the coil (bottom trace) and the disturbance created in the power grid (middle trace) are presented in Figure 10.

TABLE IV.    EM GLITCHING RESULTS FOR SINGLE SUPPLY MAX 10

| Device | Glitch parameters | | | |
|---|---|---|---|---|
| | 190V 27ns | 220V 30ns | 260V 35ns | 290V 40ns |
| 10M16SCE144 | 26 | 80 | 184 | 352 |

TABLE V.    EM GLITCHING RESULTS FOR DUAL SUPPLY MAX 10

| Device | Glitch parameters | | | |
|---|---|---|---|---|
| | 170V 31ns | 200V 34ns | 240V 30ns | 285V 30ns |
| 10M16DAF256 | 241 | 650 | 191 | 254 |

The JTAG command space was analysed for any undocumented commands. All known commands were sourced from Configuration User Guide, Boundary Scan Guide, BSDL files and *.jam files. The JTAG interface was scanned for the length of DR registers. Also, power analysis was used to spot any activities caused by undocumented commands. The following undocumented commands were found: 008, 015, 090, 091, 1EE, 206, 207, 2B0, 2D0, 303, 3F5.

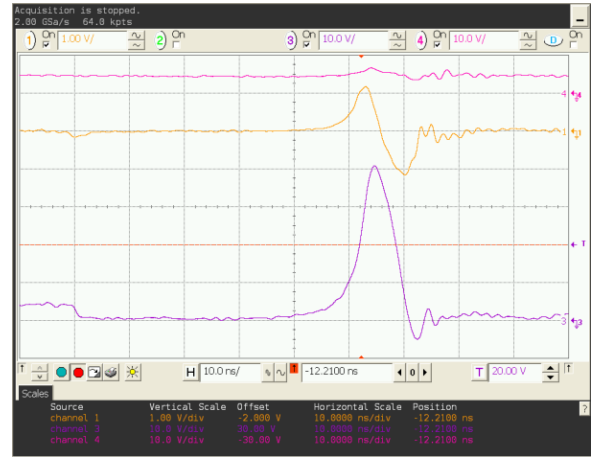

Fig. 10. Example of a successful electromagnetic glitch.

Another security vulnerability found was related to write protection of certain Flash memory regions (UFM and CFM). It was possible to change any bit in Flash memory except the system area from 1 to 0. This could be used to extract encrypted bitstream, because single bit change results in noticeable changes in the power trace.

Once the JTAG protocol was learned and implemented on the test board it was possible to scan the whole Flash memory. It was found that it has 4 regions in 10M08SCE144 device:

System area: 00000 – 007FF not readable, writable once after erase
User Flash memory: 00800 – 1CFFF always readable and writable
Configuration memory: 1D000 – 4E7FF read protectable, always writable
Shadow memory: 4E800 – 4EFFF always readable, write protected

Data remanence testing was also applied to the device. This was achieved by earlier termination of the chip erase operation. This revealed susceptibility of MAX 10 devices to these attacks. More than 97% of configuration data were successfully extracted with Verify Protect fuse enabled. It might be possible to optimise the process to achieve better success rate.

*B. Semi-invasive attacks*

Full chip laser fault injection scanning of 10M08SCE144 device with Verify Protect fuse set was performed. The whole die was scanned with the size of 4300×4400 microns. The result is presented in Figure 11. The light blue area corresponds to fault in JTAG operation, dark blue area to corrupted UFM data and red area to disabled Verify Protect fuse.

The area where unlocking of Verify Protect fuse takes place corresponds to the Flash array. In order to find the precise locations for fault injection the scanning area was reduced to 1000×1000 microns. The fault injection timing was optimised according to the information gathered from the power trace. That is the time when the power consumption is increased due to access to the Flash. The result is presented in Figure 12. The red area corresponds to the disabled Verify Protect fuse.

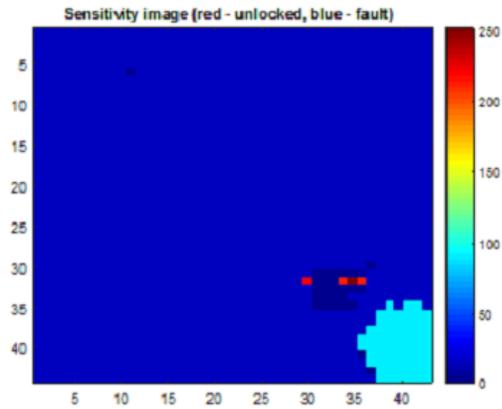

Fig. 11. Full chip laser fault injection scanning.

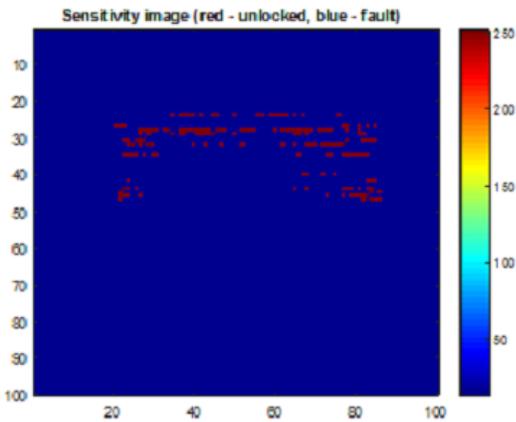

Fig. 12. Flash area laser fault injection scanning in Verify Protect mode.

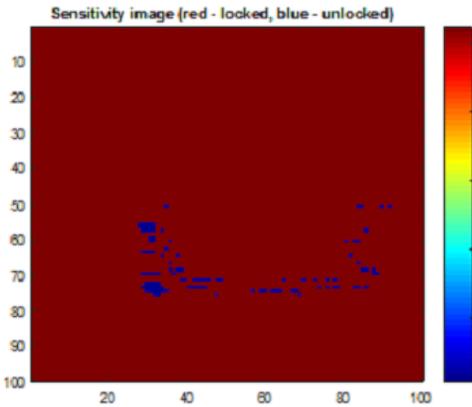

Fig. 13. Flash area laser fault injection scanning in JTAG Secure mode.

The same Flash array area was scanned on 10M04SCE144 device with JTAG Security fuse activated. In this secure mode the chip does not respond to any JTAG commands except boundary scan related. This is the most secure mode available in MAX 10 devices. The result is presented in Figure 13. The blue area corresponds to the disabled JTAG Security fuse.

In order to be more successful with power glitching attacks the timing restrictions of the data fetch from Flash memory needs to be better understood. Laser fault injection attacks can help a lot with that. To achieve this the laser was focused at one of the sensitive locations that caused both unlocking the JTAG Secure fuse and corruption of the Flash data. Figure 14 shows the results of overshooting beyond TCK rising edge. The X coordinate corresponds to the laser switched time beyond TCK in microseconds and Y coordinate to the time before TCK. Figure 15 shows the undershooting results with X coordinate corresponding to the time between switching off the laser and TCK change in 1/10 of microseconds and Y coordinate to the laser pulse time in microseconds.

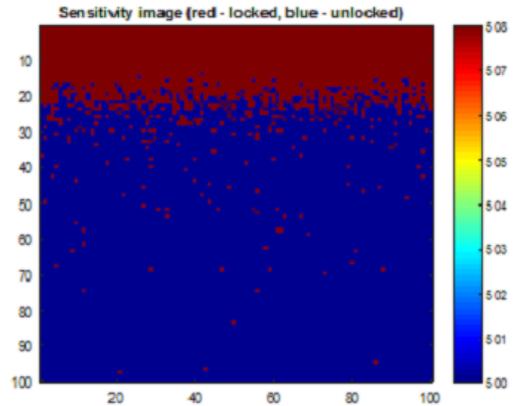

Fig. 14. Fault timing results for overshooting TCK pulse.

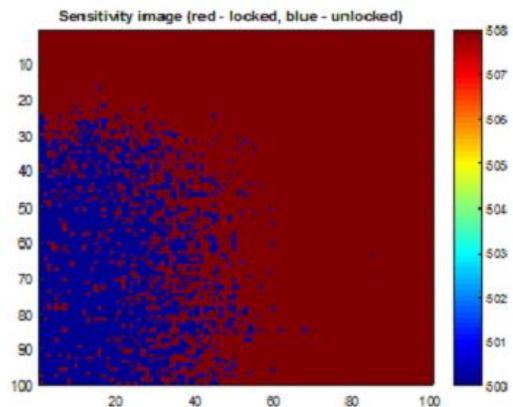

Fig. 15. Fault timing results for undershooting TCK pulse.

It can be observed that the laser pulse timing has no effect after TCK change. Also, the laser pulse time must be quite large compared to the data rate – at least 15 microseconds versus 800ns data time. This could be caused by the fact that the data from a whole row in Flash array are first loaded into a buffer before being transferred via data bus.

## C. Invasive attacks

For invasive sample preparation of 10M16SAU169 device only backside approach was used. The silicon substrate was mechanically thinned to 10 micrometers before the remaining silicon was chemically etched exposing the internal layers. Figure 16 shows the JTAG logic area under optical microscope. The transistors density is quite high and typical to that of 55nm process used in chip fabrication. The number of gates is approximately 60'000 which will make reverse engineering process quite challenging. Also all the imaging will have to be done using SEM.

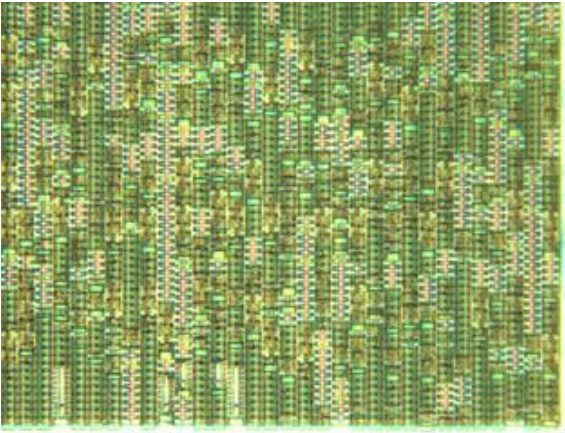

Fig. 16. JTAG area under optical microscope with 160× objective.

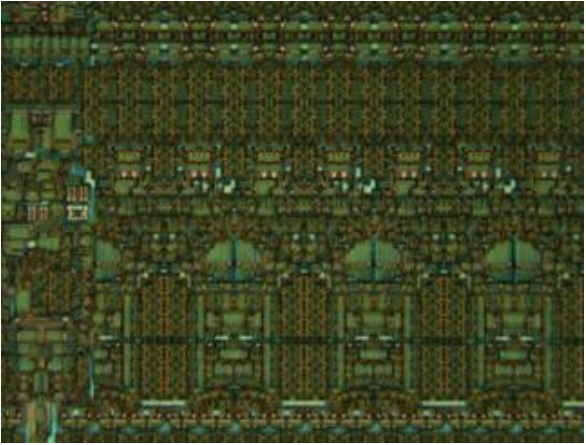

Fig. 17. FPGA logic area under optical microscope with 160× objective.

Figure 17 shows the FPGA logic area under optical microscope. Although the overall area is very large, there are many identical blocks. In order to understand the bitstream encoding the whole area of FPGA fabric will have to be reverse engineered. This will be time consuming and expensive process. Figure 18 shows the Flash array area under optical microscope. The Flash memory cells have surprisingly large size of 280×660nm and therefore are visible under optical microscope. 10M16SAU169 chip has two Flash arrays on the die. Each array is 64 bit wide with 32 columns per bit and 145×4×2 rows of data.

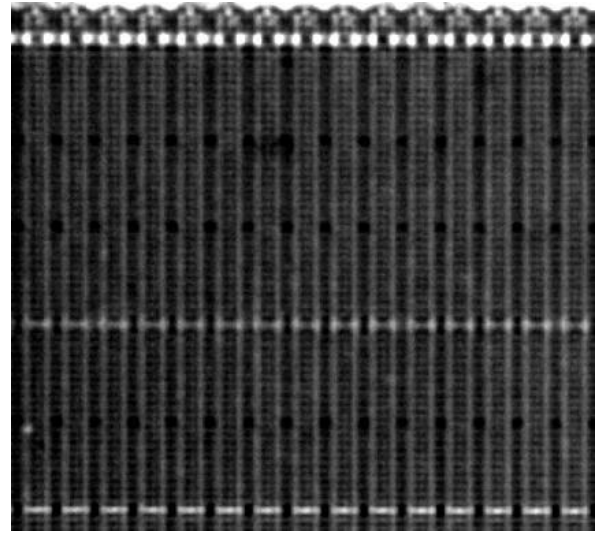

Fig. 18. Flash area under optical microscope with 160× objective.

## V. DISCUSSION

This paper outlines only some initial feasibility study research carried out on Intel MAX 10 FPGAs. Some areas will have to be investigated more thoroughly to confirm possible attacks as proof of concept.

There is a lot of security via obscurity in MAX 10 FPGAs. The lack of information about JTAG interface makes any analysis and attacks very challenging. Although many undocumented commands were found, making any use of them would require further research. Especially if someone wants to find any deliberately inserted backdoors.

With the unknown bitstream encoding format it would be very challenging to reverse engineer any design even if the bitstream is not protected. Very likely that development tools will have to be reverse engineered to understand the bitstream structure and encoding format.

Power analysis helps a lot in understanding the internal functionality of MAX 10 devices. It should be possible to extract the AES key and encrypted bitstream using DPA methods.

There are three security related fuses in MAX 10 devices: Verify Protect, Encrypted POF Only, and JTAG Security. Verify Protect and Encrypted POF Only fuses do not provide any security on their own. It is only the combination of both fuses that protects the bitstream, however, the user Flash memory (UFM) is not protected and can always be read and programmed. Moreover, even when the configuration Flash memory (CFM) is protected it can still be programmed, thus making modification attacks possible. Also, data remanence can be used to recover Flash data after full chip erase operation. It is only the JTAG Security fuse that can completely disable the access to Flash via JTAG. However, semi-invasive attacks were very successful in bypassing this protection. All these attacks are relatively easy to implement thanks to insecure implementation of the security fuses. In particular, the specific value 0x6C48A50F in the system area of the Flash which activates the security fuse.

Power glitching and EM fault injection attacks are likely to be the most dangerous if reliable and controllable way of their implementation is found. Using these attacks it might be possible to bypass the security protection without removing the chip from original board. This is especially important for devices in BGA packages, because they are harder to mount for semi-invasive attacks.

Invasive attacks can be used to recover information from Flash and to reverse engineer the JTAG circuitry. That way it would be possible to understand all undocumented JTAG commands and check for any backdoors and Trojans.

## VI. FUTURE WORK

Further analysis of MAX10 devices needs to be carried out to evaluate its security. This would involve power glitching and EM fault injection experiments with more precise timing control. Also, undocumented commands should be looked at with further brute force searches. Bitstream encryption and encoding will be another area for further investigations. Finally, attempts to reverse engineer the JTAG logic should be considered in a hope to find some backdoors or hardware Trojans.

## VII. CONCLUSION

The main purpose of this paper was to evaluate the hardware security of Intel MAX 10 low-end FPGA devices against widely known attacks. Although not specifically designed for high-security applications these devices are positioned as being highly secure. This is reflected in their datasheets with the description of various security features. This paper demonstrates that many of those security features are not robust enough.

The research described in this paper has found some serious security issues in the Intel MAX 10 devices. In many aspects the security of these devices is implemented via obscurity. The lack of documentation on the JTAG commands and AES mode of operation makes certain attacks more challenging. This is only a feasibility study paper that is aimed at outlining some weaknesses in the hardware security of MAX 10 devices. Some attacks, in particular non-invasive, will require further development work to become fully successful. However, the initial observations and preliminary results were very encouraging.

The security fuse settings are not properly documented. Verify Protect fuse only protects the configuration Flash memory (CFM) but leaves user Flash memory (UFM) fully accessible. Moreover, the access to the Flash memory is still possible via FPGA design using SRAM configuration. Encrypted POF Only fuse on its own does not protect JTAG access to the Flash memory. It is only the combination of both Verify Protect and Encrypted POF Only fuses that gives protection of configuration memory against reading. However, the user Flash memory is still readable. Moreover, the write access to both user Flash and configuration Flash is still possible. This can be used for modification attacks, for example, to extract the encrypted bitstream.

Scanning of the JTAG interface command space revealed many undocumented commands. However, very likely that silicon reverse engineering is the only reliable method for understanding functionality of those commands.

MAX 10 devices leak a lot of information via power consumption. Power analysis helps in mounting fault injection attacks by allowing more precise synchronisation. AES decryption always leaves distinctive power traces clearly distinguishable for different keys and different data. In combination with Flash modification attacks this can be used for encrypted bitstream extraction.

Further work is required to find reliable implementation of non-invasive power glitching and EM fault injection attacks. Not only they are much faster to implement (minutes vs hours for semi-invasive) but they do not require the device to be taken off the original board. This is especially convenient for devices in BGA packages.

Semi-invasive attacks in the form of laser fault injection were found to be capable of bypassing all security protection fuses in MAX 10 devices. Multiple attack points were found within the embedded Flash array. Also these attacks were found not to be very sensitive to the timing of the laser pulse. The sample preparation time for devices in TQFP packages was relatively short of just a few minutes. However, BGA packages are likely to require precise milling to prevent damaging of the power supply grid located under silicon die inside BGA carrier PCB.

For successful reverse engineering of bitstream some reverse engineering of Quartus tools will be required.

Finally, the reverse engineering of JTAG logic might lead to finding a backdoor or Trojan that would allow easy access to Flash memory including system area with AES key.

## ACKNOWLEDGMENT

This research was carried out independently from any funding body.

Intel was notified about the existence of vulnerabilities in MAX 10 devices via secure@intel.com.